\documentclass[amsmath,amssymb,showkeys]{revtex4}
\bibpunct{(}{)}{:}{a}{,}{,}

\usepackage{graphicx}
\usepackage{dcolumn}
\usepackage{bm}
\usepackage{natbib}

\newcommand {\be}{\begin{equation}} 
\newcommand {\ee}{\end{equation}}

\def \be{\begin{equation}}
\def \ee{\end{equation}}
\def \bea{\begin{eqnarray}}
\def \eea{\end{eqnarray}}

\def \kf{k_{\mbox{\tiny F}}}

\def \Del{\Delta}

\def \Qdag{Q^\dag}

\def \Qdag{Q^{\neq}}

\def \Tg{T_{\mbox{\tiny G}}}

\def \Ta{T_{\mbox{\tiny A}}}

\def \kB{k_{\mbox{\tiny B}}}

\def \a{\alpha}
\def \b{\beta}

\def \d{\delta}

\def \kf{k_{\mbox{\tiny F}}}

\def \Ea{E_{\mbox{\tiny A}}}

\def \DGfd{\Del G_{\mbox{\tiny F}\ddag}}
\def \Gdag{G_{\ddag}}
\def \Gf{G_{\mbox{\tiny F}}}
\def \mfd{m_{\mbox{\tiny F}\ddag}}
\def \DGud{\Del G_{\mbox{\tiny U}\ddag}}
\def \Gu{G_{\mbox{\tiny U}}}
\def \mud{m_{\mbox{\tiny U}\ddag}}
\def \DG{\Del G}
\def \DGfu{\Del G_{\mbox{\tiny FU}}}
\def \DH{\Del H}
\def \DS{\Del S}
\def \DCp{\Del C_{\mbox{\tiny P}}}
\def \DHud{\Del H_{\mbox{\tiny U}\ddag}}
\def \DSud{\Del S_{\mbox{\tiny U}\ddag}}
\def \DCpud{\Del C_{\mbox{\tiny P U}\ddag}}
\def \To{T_o}
\def \co{c_o}
\def \Sdag{S^{\neq}}

\begin{document}

\title{Determination of barrier heights and prefactors from protein
  folding rate data } 

\author{S. S. Plotkin$^{\ddag}$} 
\email{steve@physics.ubc.ca}
 
\affiliation{${}^\ddag$ Department of Physics and Astronomy, 
University of British Columbia, 
Vancouver, BC V6T-1Z1, 
Canada\\
}

%\date{\today}

%\newpage

\begin{abstract}
We determine both barrier heights and prefactors for protein folding
by applying constraints determined from experimental rate measurements
to a Kramers theory for folding rate.  The theoretical values are
required to match the experimental values at two conditions of
temperature and denaturant that induce the same stability.  Several
expressions for the prefactor in the Kramers rate equation are
examined: a random energy approximation, a correlated energy
approximation, and an approximation using a single Arrhenius
activation energy.  Barriers and prefactors are generally found to be
large as a
result of implementing this recipe, i.e. the folding landscape is
  cooperative and smooth. Interestingly, a prefactor with a
  single Arrhenius activation energy admits 
no formal solution.
\end{abstract}

  %Use showkeys class
				%option if keyword 

\maketitle
%\large
%\baselineskip=0.4cm

\newcounter{saveeqn}
\newcommand{\alpheqn}{\setcounter{saveeqn}{\value{equation}}
\stepcounter{saveeqn}\setcounter{equation}{0}
\renewcommand{\theequation}
      {\mbox{\thesection\arabic{saveeqn}\alph{equation}}}}
\newcommand{\reseteqn}{\setcounter{equation}{\value{saveeqn}}
\renewcommand{\theequation}{\thesection\arabic{equation}}}

%\newpage
\section{Introduction}

In contrast to many exothermic reactions in organic chemistry,
% such as enolate
%formation, 
the log protein folding rate displays a significant linear trend with the relative
stability of the product and reactant (folded and unfolded states)~\citep{FershtAR99:book}. This indicates a
late transition state in the language of Hammond's postulate, and
the slope of the log rate vs. stability line  quantifies the degree of native
structural information in the transition state.

Native stability may be  modified by adjusting temperature $T$ or denaturant concentration $c$.
Many proteins show  linearity over the majority of the branches of
their Chevron plot, implying  a linear dependence of folding and unfolding barriers on
denaturant concentration $c$~\citep{JacksonSE91i}: 
\alpheqn
\begin{eqnarray}
\DGfd (T,c) \equiv \Gdag(T,c) - \Gf(T,c) = \DGfd (T,0) - \mfd c 
\label{eq:1}
\\
\DGud (T,c) \equiv \Gdag(T,c) - \Gu(T,c) = \DGud (T,0) + \mud c
\label{eq:2}
\end{eqnarray}
\reseteqn
with $\mfd >0$ and $\mud >0$. 

Subtracting~\ref{eq:2} from~\ref{eq:1}, and defining $\DG \equiv
\DGfu = \Gu -\Gf$ and $m = \mfd + \mud$ we have that
\be
\DG (T,c) = \DG (T,0) - m c
\label{eq:dg}
\ee
For 2-state folders the kinetically determined $m$ above equals to
good approximation the
thermodynamically determined $m$-value from relative stabilities.

Applying Kramers rate theory, the log forward folding rate is given by 
\bea
\ln \kf (T,c) &=& \ln k_o (T,c) - \DGud (T,c)/T \nonumber \\
&=& \ln k_o (T,c) -  (\DGud (T,0) + \mud c )/T
\label{eqlnko}
\eea

Eliminating $c$ from equations~\ref{eq:dg} and~\ref{eqlnko} gives
\be
\ln \kf (T,c) - \frac{\mud}{m} \frac{\DG (T,c)}{T} = \ln k_o (T,c) - \frac{1}{T}
\left( \DGud (T,0) + \frac{\mud}{m} \DG (T,0) \right) \: ,
\label{eq:rate}
\ee
where the left hand side of~\ref{eq:rate} depends on both $(T,c)$, but the function on the
right hand side depends on $c$ only through the prefactor.
Empirically it was
observed by Scalley et al~\citep{ScalleyM97} that for the proteins
CspB and protein L, 
the data for various $c$ collapse
onto a single curve when the left hand side is plotted vs. $1/T$. This
indicates that the right hand side is a function of temperature alone and so $\ln k_o
(T,c) \approx \ln k_o (T)$. 
Denaturant
concentration does not have a significant effect on the rate the
system escapes from local traps (at least for those proteins studied). 
We make this assumption here as well. 

Because the prefactor is independent of $c$, the change in log folding
rate with denaturant is directly proportional to the change in barrier
with denaturant:
\be
\d \ln \kf \equiv \ln \kf (T,c) - \ln \kf (T,0) =
- \left( \DGud (T,c) - \DGud (T,0) \right)/T
= - \d \DGud /T
\ee
which together with eq.~\ref{eq:rate} gives 
\bea
\d \DGud &=& - (\mud/m) \d \DG 
\label{dgdg} \\
\d \ln \kf &=& (\mud/m) (\d \DG /T) \: .
\label{ratestab}
\eea
This quantifies the assertion above that log folding rates depend
linearly on the relative stability of the products. If we let 
$\mud/m \equiv \Qdag$, eq.~\ref{dgdg} can be rewritten
as 
\be
\d \Gdag = \Qdag \d \Gf + (1-\Qdag) \d \Gu
\label{eqlinear}
\ee
which is the commonly used linear free energy relation~\citep{BryngelsonJD95}.

Inspection of rate-stability isotherms for several different proteins
(cytochrome C~\citep{Mines96}, protein L~\citep{ScalleyM97},
cspB~\citep{SchindlerT96}, N-terminal protein L9~\citep{KuhlmanB98}, S6~\citep{OtzenDE04}) shows
linearity over ranges up to $\approx 25 
\mbox{kJ} \cdot \mbox{mol}^{-1} \approx 10 \kB T$, indicating large and
robust folding barriers- substantially larger than the folding barriers seen
in many simulations for example (c.f. figure~\ref{fig:gray}).

At a higher temperature, the log rate vs. stability curve is still
linear, with approximately the same slope, indicating the nativeness
of the transition state, in terms of solvent exposure, is not
significantly changed (Fig.~\ref{fig:gray}). However the rates are higher,
presumably due to 2 effects: 1.) the prefactor increases at higher
temperature (since activated escape from traps is further facilitated,
and solvent viscosity is reduced),
and 2.) the thermodynamic weight of the entropic component to
the barrier (which includes contributions from the solvent) increases
as well, which may decrease the barrier height.

\section{Methodology}

In what follows we apply Kramers rate theory together with energy
landscape ideas to extract barrier heights and prefactors
from experimental rate data. 

The temperature dependence of the stability is given by the
Gibbs-Helmholtz expression~\citep{FershtAR99:book,JacksonSE91i}:
\be
\DG (T,c) =  \DH  - T \DS  + \DCp \left( T- T_o - T \ln \left(
T/T_o\right) \right) \: .
\ee
Then at equal stabilities $\DG (T_o,c_o) = \DG (T,c)$:
\be
\DCp \left( T- T_o - T \ln \left(
T/T_o\right) \right) = \left( T - T_o
\right) \DS + m \left( c - c_o \right) \: .
\label{eqstabCp}
\ee

For two-state folders, the heat capacity ratio $\DCpud/\DCp$ is
approximately equal to  m-value ratio $-\mud/m$, giving 
the fractional solvent accessibility of the transition state. 
We assume this equality here as well, which gives for
eq.~\ref{eqstabCp}:
\be
\DCpud \left( T- T_o - T \ln \left(
T/T_o\right) \right) =  - \frac{\mud}{m} \DS \left( T - T_o
\right)  - \mud \left( c - c_o \right) \: .
\label{eqstabCp2}
\ee

Inserting eq.~\ref{eqstabCp2} into Gibbs-Helmholtz expressions for
the barrier heights $\DGud$ at $(T,c)$ and $(T_o,c_o)$ gives the change in
barrier height {\it at fixed stability}:
\be
\left[\DGud (T,c) - \DGud (T_o, c_o)\right]_{\DG(T,c)=\DG(T_o,c_o)}
\equiv \d'\DGud  =
-\left( T- T_o\right) \left( \DSud + \frac{\mud}{m} \DS \right) \: ,
\label{eq:dDG}
\ee
which is independent of $c$ and depends only on the temperature difference between the two
fixed-stability states (and thermodynamic parameters). This equation
applies to points A and B in 
figure~\ref{fig:gray} for example.

For changes in temperature of a few degrees, the change in barrier
height $\d'\DGud$ is only a few percent of the total barrier height,
when the rates vs. temperature and denaturant are fit to a
model to extract thermodynamic parameters, as in
ref.s~\citep{ScalleyM97,SchindlerT96,KuhlmanB98,OtzenDE04}. We used
equation~\ref{eq:dDG} for the change in barrier height when
thermodynamic data were available. For the case of cytC we set $\d' \DG =0$.

The rates for pairs of states at the same stability $\DG$ are given from eq.~\ref{eqlnko} as
\alpheqn
\begin{eqnarray}
\ln \kf (T_o,c_o) &=& \ln k_o (T_o) - \DGud (T_o,c_o)/T_o 
\label{eq:k1}
\\
\ln \kf (T,c) &=& \ln k_o (T) - \DGud (\To,\co)/T - \d'\DGud  /T
\label{eq:k2}
\end{eqnarray}
\reseteqn

\subsection{Random energy model for the temperature-dependent prefactor}

At the mean field level for a landscape of uncorrelated
states (Random energy model or REM), the temperature-dependence of the 
prefactor in 
equation~\ref{eqlnko} is
super-Arrhenius~\citep{BryngelsonJD89,Onuchic97}. Moreover the prefactor goes as
the reciprocal of the viscous friction
coefficient~\citep{Hanggi90,SocciND96:jcp,Klimov97}, so the log prefactors 
at  $(T_o, c_o)$ and $(T,c)$ may be written as
\alpheqn
\bea
\ln k_o (T_o) &=& \ln k_{oo} - \Del^2/2 T_o^2
\label{pref0}
\\
\ln k_o (T) &=& \ln k_{oo} - \Del^2/2 T^2 + \ln\left(
\eta(T_o)/\eta(T)\right) \: .
\label{pref1}
\eea
\reseteqn
To compare rate theories with experimental data we must introduce a
fundamental time scale or rate constant $k_{oo}$, which is then
modified by barriers representing the ruggedness of the energy
landscape. Rates for short loop closure are about 
$2 \times 10^7 \mbox{s}^{-1}$~\citep{LapidusLJ00}, comparable to helix
formation rates of $\sim 10^7 \mbox{s}^{-1}$, and somewhat faster than
rates of hairpin formation $\sim 10^6
\mbox{s}^{-1}$~\citep{EatonWA00}. We take $10^7 \mbox{s}^{-1}$ as an
estimate of the fastest local rate. Since $\sim 10 -100$ loops and/or
secondary structural elements exist in a protein, we then take $k_{oo}
= 10^9 \mbox{s}^{-1}$. We will see later that larger estimates for
$k_{oo}$ give larger estimates for inferred folding barriers.
We use the known temperature dependence of the
viscosity in water~\citep{CRC}. 
The quantity $\Del^2$
measures the ruggedness of the
energy landscape. It may be eliminated from~\ref{pref0}
and~\ref{pref1} to give an equation relating the prefactors:
\be
\ln k_o (T) = \left(1-\frac{T_o^2}{T^2}\right) \ln k_{oo} +
\frac{T_o^2}{T^2} \ln k_o (T_o) +
\ln\left(\frac{\eta(T_o)}{\eta(T)}\right) \: .
\label{bothpref}
\ee

Equations~\ref{eq:k1}, \ref{eq:k2}, and~\ref{bothpref}
constitute a system of 3 linear equations for 3 unknowns: $\DGud (\To,\co)$, $\ln
k_o (T_o)$ and $\ln k_o (T)$, which can be solved analytically at any
given stability, from linear fits to the log rate-stability data.

%\subsection{Simulation model}

%Notes on mapping c to change in $\En$

\section{Results}

The results of applying the method are shown in figure~\ref{fig:grayDG}, for the data in
figure~\ref{fig:gray}, ranging from the stability of wild type at
$296K$ ($-74 \mbox{kJ/mol}$) to zero stability at the transition
midpoint.  Barrier heights are plotted in units of kJ/mol, rates in
prefactors are in units of $\mbox{s}^{-1}$.

We can see several things from this plot. The barrier heights at the
transition midpoint are large, compared to values obtained from
simulation models as well as theories with pair interaction
potentials. If the linear relation in
eq.~\ref{dgdg} held until the transition midpoint, the barrier
would be about $30$ kJ/mol plus whatever the barrier was at conditions
of zero denaturant.

The slope $\d \DGud/\d \DG \approx 0.8 $ is also larger
than its empirical value of $\mud/m \approx 0.4$~\citep{Mines96}, thus the barriers vanish
at weaker stabilities than the wild type protein. This indicates a
breakdown in the validity of the theory at higher stabilities (larger $\DG$).

There are 2  parameters in the theory for which we have put in
approximate values: the value of the attempt frequency $k_{oo} =
10^{9} \mbox{s}^{-1}$, and the value of $\d'\DGud$, which we have set
to zero for cytochrome~C in the absence of an empirically determined value. Increasing
$k_{oo}$ or decreasing $\d'\DGud$ raises barriers, but does not change
the slope $\d \DGud/\d \DG$. The value of $-\DG$ where the barrier
vanishes linearly decreases as $\d'\DGud$ is decreased below zero,
with the barrier vanishing at the stability of the wild type when
$\d'\DGud$ is about $-1.6$ kJ/mol. This is not an unreasonable number
compared to experimental numbers for other proteins (see below), however it is
somewhat disconcerting that barrier heights are such a strong function
of the barrier change $\d'\DGud$. We will see later that this
sensitivity is not present when a correlated landscape model is used for
the prefactor. 

Figure~\ref{fig:grayDG} also shows that at least for the REM approximation it
is important to account for changes in the viscosity of the solution
with temperature, as the barrier substantially decreases when the
viscosity is held constant vs. temperature. 

Equation~\ref{pref0} or~\ref{pref1} may now be solved for $\Del$, giving a number
$\approx 15$ kJ/mol, that only weakly depends on stability $\DG$ or
barrier change $\d'\DGud$.
Estimating the chain conformational entropy as $\sim 100 \kB$~\citep{DaquinoJA96,LeachSJ66}, we can
give an estimate for the glass temperature $\Tg$ for this system, 
\be
\Tg = \Del/(2 S_o/\kB)^{1/2}
\label{tg}
\ee
which is also a fairly robust
number as a function of stability or barrier change, as shown in 
figure~\ref{fig:tg}. At the stability of wild type cyt C, $\Tg \approx
150K$, giving $T/\Tg \approx 2.0$ at $296$K.

\subsection{Correlated landscape model for the temperature-dependent prefactor}

Many of the problems of the REM approximation are resolved by accounting
for pair correlations between states in the expression for the
prefactor. Below a critical temperature
$\Ta$ on a correlated landscape, dynamics are activated, and the rate prefactor increases as
temperature is raised~\citep{WangPlot97,PlotkinSS02:quartrev1,PlotkinSS02:quartrev2}. 
The expressions for the rate prefactors at $T_o$ and $T$ become
\alpheqn
\bea
\ln k_o (T_o) &=& \ln k_{oo} - (\Sdag/2) \left(\a - \b \left( 1 - \Tg/\To
\right)^2 \right) 
\label{lnkogrem} \\
\ln k_o (T) &=& \ln k_{oo} - (\Sdag/2) \left(\a - \b \left( 1 - \Tg/T
\right)^2 \right) + \ln \left( \eta(\To)/\eta(T) \right) \: .
\label{lnkgrem}
\eea
\reseteqn
Here $\Sdag$ is the chain entropy at the transition state, and $\a$
and $\b$ are parameters measuring the mismatch between entropy and
energy giving the typical free energy barrier governing trap escape. The 
values for a bulk polymer $\a \approx 0.5$, $\b \approx 1.8$ are used
below~\citep{WangPlot97,PlotkinSS02:quartrev1,PlotkinSS02:quartrev2}.
The temperature $\Tg$ was adjusted to the value  that reproduced
the experimentally determined slope of barriers vs. stability,
$\mud/m$. In table~I this number is compared to the value of $\Tg$ that
emerges from the REM analysis. A mismatch of these 2 values may
indicate a breakdown of the REM approximation for states in
determining prefactors, i.e. a breakdown in the validity of
eq.s~\ref{pref0},b. 
For cyt C the value of $\Tg$ giving the correct slope is about $1.2$
kJ/mol, vs. $1.0$  kJ/mol from the REM analysis.

The entropy may be eliminated from~\ref{lnkogrem}
and~\ref{lnkgrem}, giving an equation that relates the prefactors,
and replacing eq.~\ref{bothpref}:
\be
\left[\ln k_{oo} - \ln k_o (T)+ \ln \left( \eta(\To)/\eta(T)
  \right)\right]
\left[\a - \b \left( 1 - \Tg/\To\right)^2 \right] =
\left[\ln k_{oo} - \ln k_o (\To)\right]\left[\a - \b \left( 1 -
  \Tg/T\right)^2 \right] 
\label{eqpref2}
\ee

Equations~\ref{eq:k1}, \ref{eq:k2}, and~\ref{eqpref2} again define 
a system of 3 linear equations for 3 unknowns: $\DGud (\To,\co)$, $\ln
k_o (T_o)$ and $\ln k_o (T)$, which may be solved analytically.
The results are shown in figure~\ref{fig:graygrem}. 

We see that both barriers and prefactors are larger than the
corresponding REM values, and the analysis for other proteins yields
quite large numbers in general (c.f. table~I for numbers). The barriers at
the transition midpoint are about $22 \kB T_{300K}$, and prefactors
are almost unactivated. 
%Moreover the slope $\d \DGud/\d \DG$ is precisely the experimental
%value of $\mud/m =0.4$, without adjusting any parameters. 
%This
%situation was fortunate- in general the slope
%obtained with this method tended to differ from the experimental
%slope, and $\Tg$ was adjusted to reproduce the experimentally
%determined value, $\mud/m$. 
The REM value of $\Tg$ resulted from approximating a value of $100
\kB$ for the
chain entropy $S_o$, so it is feasible that this estimate for the REM $\Tg$ could differ
from the $\Tg$ that gives the correct $\mud/m$.
The parameters $\a$ and $\b$ could in principle have been adjusted to best match the
experimental slope, however it can be shown that this results in the
same solution of~\ref{eq:k1}, \ref{eq:k2}, and~\ref{eqpref2} as that
determined by varying $\Tg$. 

In contrast to the REM approximation, the effects of the temperature dependence of
viscosity were not significant here (figure~\ref{fig:graygrem}). Nor
were there any significant effects due to barrier height difference- as
$\d' \DGud$ changed from $-2$ kJ/mol to $0$ kJ/mol, the barrier
changed by less than $2\%$. The effects due to $\Tg$ are modest as well:
over the range of $\Tg$ values in figure~\ref{fig:tg}B, the barrier
height changed by less than $15\%$. Lastly, the prefactors of the
correlated landscape model are nearly constant over the range of
experimental stabilities (figure~\ref{fig:graygrem}), consistent with
empirical observations (c.f. the comments below eq.~\ref{eq:rate}).

Equation~\ref{lnkogrem} or~\ref{lnkgrem} may now be solved for $\Sdag$ as a
check, giving $\Sdag \approx
40 \kB$, or about $40\%$ of the unfolded chain entropy assumed
in finding the REM $\Tg$. 
Alternatively we can estimate the unfolded entropy $S_o$ from the
value of $\Sdag$ as $\Sdag \approx (1-\mud/m) S_o$, then eq.~\ref{tg} gives
$\Del \approx 14$ kJ/mol. Since the variances of individual residues
add to give $\Del^2$, $\Del^2 \approx N (1-\mud/m) b^2$, where $b$ is
a non-native energy scale per residue, here $\approx 0.7 \kB
T_{300}$. 

Figure~\ref{lnkoofig} shows that the inferred barriers and prefactors increase as
the value of the bare reconfiguration rate $k_{oo}$ increases. The prefactor $\ln k_o
(T_o)$ closely follows the bare reconfiguration rate $\ln k_{oo}$,
i.e. they are roughly equal. The barriers at the
transition midpoint $\DGud^o$ and at the stability of the wild-type
protein $\DGud^{(wt)}$ increase linearly, as $\sim 2 T_o \ln k_{oo}$. 
%For $k_{oo} = 10^9 \mbox{s}^{-1}$, $\DGud^o \approx 22 \kB T_{300}$, 
%$\DGud^{(wt)} \approx 10 \kB T_{300}$, and $k_o (T_o) \approx 4 \times
%10^8 \mbox{s}^{-1}$.

In the REM analysis there is an intermediate regime where the prefactor
has  a more complex
temperature dependence than eq.~\ref{pref0}. We do not describe this
regime in detail since it is obtained from eq.s~\ref{lnkogrem}
and~\ref{lnkgrem} in the limit that $\a\rightarrow 1$, $\b \rightarrow
2$, $\Sdag \rightarrow S_o$. Values obtained tended to be bracketed by the
REM and correlated models.

For NTL9, the solution of the REM gave a $\Tg$ that
monotonically decreased from a value of $0.4$ at the stability of
the wild-type protein, to zero at a stability of about $11$
kJ/mol. Similarly the prefactor monotonically increases from $10^8
\mbox{s}^{-1}$ at the stability of the wild-type to $10^{10}
\mbox{s}^{-1}$ at zero stability. We note that these problems are
not present if the stability difference $\d'\DGud$ is set to zero,
if the prefactor is $2$ or more orders of magnitude slower, or if
the temperature-dependence of the viscosity is neglected. We take this sensitivity as a
shortcoming of the procedure of rigorously demanding that the
landscape theory fit to a limited subset of the experimental
data. In this sense a best (but not exact) fit to experimental rate
surfaces as a function of both $T$ and $c$ as
in~\citep{SchindlerT96,ScalleyM97,KuhlmanB98,OtzenDE04} is likely to give more
accurate numbers. Likewise in the correlated model for NTL9, the prefactor increased
from about $10^8 \mbox{s}^{-1}$ at the stability of the wild-type to
unphysical values at zero stability. A similar situation exists in the
REM recipe for protein S6, however it is resolved in the correlated
landscape model for that protein.

CspB showed some difficulties that arose from its unusually late transition state
($\mud/m \approx 0.9$)~\citep{PerlD02}. The parameter $\Tg$ in the
correlated model could not be adjusted to reproduce the
high slope of barrier vs. stability, without giving negative
barriers. Again this may be an artifact of the exact fitting
method mentioned above, i.e. more
experimental data may also be needed to obtain more accurate numbers, or it may indicate that a simple
mean field prefactor does not fully adequately describe the folding
dynamics of this protein. 
%In table~I we give the barriers and prefactors that result 
%from choosing the REM $\Tg$. At this value of $\Tg$, the prefactor
%became unphysically large- larger than the total bare reconfiguration rate of
%$10^{12}\mbox{s}^{-1}$. 
In this case we took the temperature $\Tg =1.81$ kJ/mol 
 that induced the barrier to vanish at the stability of the wild
type protein. This has a steep barrier-stability curve,
with slope $\mud/m = 0.8$ (as opposed to $0.9$
 observed empirically), very small barrier ($7$
kJ/mol at zero stability), and rugged landscape with 
very slow prefactor (about $10^2 \mbox{s}^{-1}$). Such small barriers
are consistent with estimates taken from simulations using
$C_{\a}$-models~\citep{SheaJE01}.

\subsection{The Arrhenius model generally admits no solution}

A model often proposed for the prefactor assumes an Arrhenius
temperature-dependence with single activation energy $\Ea$, so that
eq.s~\ref{pref0} and~\ref{pref1} are replaced
by 
\alpheqn
\bea
\ln k_o (T_o) &=& \ln k_{oo} - \Ea/ T_o
\label{pref0Ea}
\\
\ln k_o (T) &=& \ln k_{oo} - \Ea/T  + \ln\left(
\eta(T_o)/\eta(T)\right) \: ,
\label{pref1Ea}
\eea
\reseteqn
from which $\Ea$ may be eliminated yielding
\be
\ln k_o (T) = \left(1-T_o/T\right) \ln k_{oo} +
\left(T_o/T\right) \ln k_o (T_o) +
\ln\left(\eta(T_o)/\eta(T)\right) \: .
\label{bothpref2}
\ee
This equation relating the prefactors together with eq.s~\ref{eq:k1}
and~\ref{eq:k2} are the new system of equations to be solved. 

Eliminating $\DGud$ from~\ref{eq:k1} and~\ref{eq:k2} gives another
equation relating the prefactors:
\be
\ln k_o (T) = \ln \kf (T,c) - (T/\To) \ln \kf (T_o,c_o) + \d' \DGud/T
+ (\To/T) \ln k_o (T_o) \: .
\label{middd}
\ee
Equations~\ref{middd} and~\ref{bothpref2} both have $\ln k_o (T)$ on
the left hand side and $(\To/T) \ln k_o (T_o)$ on the
right. Subtracting them then gives an equation that is independent of
any variable to be solved for:
\be
\ln \kf (T,c) - (T/\To) \ln \kf (T_o,c_o) + \d' \DGud/T = 
\left(1-T_o/T\right) \ln k_{oo} +\ln\left(
\eta(T_o)/\eta(T)\right)
\ee
which cannot be true in general, in particular because the left  hand
side depends on $c$ and the right hand side does not. 

A geometric analog may be helpful in understanding the situation. The
solution to 3 equations in 3 variables is equivalent to finding the
point where 3 planes intersect. Letting 
\bea
x_1 &=& \ln k_o (T_o) \nonumber \\
x_2 &=& \ln k_o (T) \nonumber \\
x_3 &=& \DGud \: , \nonumber
\eea
equations~\ref{eq:k1},~\ref{eq:k2}, and~\ref{bothpref2} may be recast
as
\alpheqn
\bea
x_2 - (\To/T) x_1 &=& A \label{eqa1} \\
x_2 - (\To/T) x_1 &=& B \label{eqa2} \\
x_1 - (1/\To) x_3 &=& C \label{eqa3}
\eea
\reseteqn
where 
\bea 
A &=& (1-T_o/T) \ln k_{oo} + \ln(\eta(T_o)/\eta(T)) \nonumber \\
B &=& \ln \kf (T_o,c_o) + (T_o/T) \ln \kf (T,c) + \d' \DGud/T
\nonumber \\
C &=& \ln \kf (T_o,c_o) \: . \nonumber
\eea
Since $A \neq B$ in general, eq.s~\ref{eqa1} and~\ref{eqa2} depict two
parallel planes. Thus there is no point of intersection and the system of equations
is ill-posed. For the special case of $A=B$ there is a whole family of
solutions consistent with the rate equations, but as mentioned above
this scenario can only hold under very special circumstances.

\section{Conclusions and discussion}

We have proposed here a method of testing energy landscape theory by
mapping  Kramers rate theory, with prefactors given from the statistics
of energies of states, to experimental data on protein folding rates. 
We considered 3 models for the prefactor here: one where
ruggedness is treated with a random energy approximation, one where
correlations are taken into account, and an Arrhenius model with a
single barrier dressing reconfiguration times. 

The numerical values of the barriers obtained from the above recipes
should be taken with a  
grain of salt, however it consistently emerged that folding barriers
were large (except for CspB): the average
barrier at the transition midpoint for the REM analysis is about $19 \kB
T$, and the corresponding barriers in the correlated model is about
$18 \kB T$. If CspB is omitted the barriers are $21 \kB
T$ and $22 \kB T$ respectively. Wild-type S6, a protein known to fold
very cooperatively~\citep{LindbergM02}, had the highest barriers.

With the exception of CspB,
the prefactors in the correlated model tended to be quite
high - approximately the bare reconfiguration rate for the whole protein
($10^{9} \mbox{s}^{-1}$). The folding barrier obtained from the recipe
decreases as estimates for the bare reconfiguration rate 
decrease (Fig.~\ref{lnkoofig}). The prefactors from the REM recipe
varied considerably. 

All of the proteins analyzed here are considered 2-state folders, so
we would expect a Kramers theory to describe them. In lower
temperature regimes the distribution of first passage times may be
more relevant to
study~\citep{PlotkinSS02:quartrev2,ZhouY03}. 

We found that in practice it was quite important to have accurate fits
for the empirical rate-stability curves. For example, as temperature
increased, the slope of the log rate vs. stability curve had to remain
roughly constant or tend to increase, to obtain reasonable solutions
of the rate equations. Otherwise we found an unphysical situation
where barriers did not increase as stability decreased. This
sensitivity to the experimental data may favor a less stringent fit to
the experimental constraints.

In fact, reflection on the procedure raises a general issue on the
rigorous application
of experimental constraints to energy landscape theory. For example, if we were to add data
at a third temperature $T_1$, two new equations would be introduced
according to the recipe- one
Kramers rate equation and one landscape equation for the prefactor,
but only one new variable is introduced- the prefactor $\ln k_o
(T_1)$. The system becomes overdetermined. Demanding equality rather
than a best fit at several temperatures becomes too stringent a
constraint on the theory, as long as the parameters in the theory
(e.g. $\Del^2$ or $\Ea$) are fixed. The more temperatures used, the
more variables must be introduced into the theory, or the parameters
must themselves become temperature-dependent. 
Nevertheless, the fact that  the Arrhenius activation model fails in general to
provide a solution for even 2 temperatures (2 data points) should
probably be seen as evidence
against its strict applicability. 

A perhaps more viable method
would be to fit several temperatures with functional forms such as
equations~\ref{pref0}, \ref{lnkogrem}, or~\ref{pref0Ea} to extract
parameters such as $\Del^2$ and $\Ea$. The difficulty in previous fits
to data has been in the separation of $\Ea$ and the activation
enthalpy $\DHud$~\citep{ScalleyM97}.
One can ask which
temperature dependence ($\Ea/T$ or $\Del^2/T^2$) gives the best fit to
the data, but there is not yet enough accurate data to distinguish
between the two scenarios~\citep{ScalleyM97,KuhlmanB98} by this
method. However the Arrhenius model becomes severely restricted by
applying experimental constraints rigorously at two temperatures and
denaturant concentrations, at the same stability. Because the
activation energy in the prefactor can be absorbed into the enthalpic
part of the barrier, and only the entropic part of the barrier is
relevant in determining rate differences at fixed stability (by
eq.~(\ref{eq:dDG})), the activation energy becomes irrelevant, and the
difference in rates must then be due to quantities independent of
denaturant concentration (entropic part of the barrier,
temperature-dependent viscosity\ldots). All rate-stability curves for a given
protein must be parallel
in the Arrhenius model- a situation not observed empirically.

Topological features of the native structure have been neglected in
the rate theory. Including polymer physics into the theoretical
model~\citep{ShoemakerWang99,PlotkinSS00:pnas,PortmanJJ01:jcp} 
may also eliminate some of
the sensitivity of the theoretically derived values in table~I on
the experimental data.

Other methods have been used to estimate barrier heights. Adding a
3-body contribution to a pair-wise interacting energy function to give
best agreement with experimental $\phi$-values, a
barrier height for protein~L of about $16$
kJ/mol was obtained~\citep{EjtehadiMR04}. Other proteins such as FKBP and CI2 had
larger barriers of $25$ kJ/mol and $42$ kJ/mol
respectively~\citep{EjtehadiMR04}. 
The large barriers observed here also suggest that 
many-body interactions may be playing a significant role in the energy function.
A variational theory for the free
energy surface of $\lambda$-repressor gave a barrier of approximately
$12$ kJ/mol~\citep{PortmanJJ01:jcp}. 
All-atom simulations of a three-helix bundle fragment of protein~A in
explicit water gave barrier heights $\approx 17$ kJ/mol at the
transition midpoint~\citep{GarciaAE03:pnas}.
Applying Kramers theory with an
experimentally determined estimate for the prefactor gave an estimate
for the free energy barrier of about $18$ kJ/mol for the cold shock
protein Csp{\it Tm}~\citep{SchulerB02}. An analysis which took
prefactors from experimental data, along with a thermodynamic analysis
to extract enthalpic and entropic contributions to the barrier, gave
typical barrier heights of about $30$ kJ/mol for the proteins
analyzed~\citep{AkmalA04}. However these last two methods found barrier
heights under conditions of zero denaturant- the barrier heights at
zero stability would likely be significantly higher. For example,
the average $\left< (\mud/m) \DG \right>$ for the proteins in table~I
is about $17$ kJ/mol, to be added to the barrier height at conditions
of zero denaturant. 

Applying this method to a simulation model, where one knows the answers
in advance, provides a good control for the study and is a topic for
future work. \\

S. S. P. acknowledges support from the
Natural Sciences and Engineering Research Council and the Canada
Research Chairs program. We thank Jos\'{e} Onuchic,  Reza Ejtehadi,
Magnus Lindberg, and Matthias Huber for helpful discussions, and M. Oliveberg for
sharing unpublished data.

%\newpage

\section{References}

\pagestyle{plain}
\addcontentsline{toc}{section}{References}

%\newpage
%\bibliographystyle{nature-doc}
%\bibliographystyle{astron}
%\bibliography{/home/steve/Madonna/Tex/Bib/steve}

\begin{thebibliography}{32}
\providecommand{\natexlab}[1]{#1}

\bibitem[{Akmal and Munoz(2004)}]{AkmalA04}
Akmal, A., and V.~Munoz. 2004.
\newblock The nature of the free energy barriers to two state folding.
\newblock \emph{Proteins}. 57:142--152.

\bibitem[{Bryngelson et~al.(1995)Bryngelson, Onuchic, Socci, and
  Wolynes}]{BryngelsonJD95}
Bryngelson, J.~D., J.~N. Onuchic, N.~D. Socci, and P.~G. Wolynes. 1995.
\newblock Funnels, pathways and the energy landscape of protein folding.
\newblock \emph{Proteins}. 21:167--195.

\bibitem[{Bryngelson and Wolynes(1989)}]{BryngelsonJD89}
Bryngelson, J.~D., and P.~G. Wolynes. 1989.
\newblock Intermediates and barrier crossing in a random energy model (with
  applications to protein folding).
\newblock \emph{J Phys Chem}. 93:6902--6915.

\bibitem[{B.Schuler et~al.(2002)B.Schuler, Lipman, and Eaton}]{SchulerB02}
B.Schuler, E.~A. Lipman, and W.~A. Eaton. 2002.
\newblock Probing the free-energy surface for protein folding with
  single-molecule fluorescence spectroscopy.
\newblock \emph{Nature}. 419:743--747.

\bibitem[{CRC(2003)}]{CRC}
CRC. 2003.
\newblock \emph{In} CRC Handbook of Chemistry and Physics. D.~R. Lide, editor,
  84th Ed. CRC Press, New York.

\bibitem[{D'Aquino et~al.(1996)D'Aquino, Gomez, Hilser, Lee, Amzel, and
  Freire}]{DaquinoJA96}
D'Aquino, J.~A., J.~Gomez, V.~J. Hilser, K.~H. Lee, L.~M. Amzel, and E.~Freire.
  1996.
\newblock The magnitude of the backbone conformational entropy change in
  protein folding.
\newblock \emph{Proteins-structure Function Genetics}. 25:143--156.

\bibitem[{Eaton et~al.(2000)Eaton, Munoz, Hagen, Jas, Lapidus, Henry, and
  Hofrichter}]{EatonWA00}
Eaton, W.~A., V.~Munoz, S.~J. Hagen, G.~S. Jas, L.~J. Lapidus, E.~R. Henry, and
  J.~Hofrichter. 2000.
\newblock Fast kinetics and mechanisms in protein folding.
\newblock \emph{Annu Rev Biophys and Biomolec Struct}. 29:327--359.

\bibitem[{Ejtehadi et~al.(2004)Ejtehadi, Avall, and Plotkin}]{EjtehadiMR04}
Ejtehadi, M.~R., S.~P. Avall, and S.~S. Plotkin. 2004.
\newblock Three-body interactions improve the prediction of rate and mechanism
  in protein folding models.
\newblock \emph{Proc Nat Acad Sci USA}. 101:15088--15093.

\bibitem[{Fersht(1999)}]{FershtAR99:book}
Fersht, A.~R. 1999.
\newblock Structure and mechanism in protein science, 1st Ed. W. H. Freeman and
  Co., New York.

\bibitem[{Garcia and Onuchic(2003)}]{GarciaAE03:pnas}
Garcia, A.~E., and J.~N. Onuchic. 2003.
\newblock {Folding a protein in a computer: An atomic description of the
  folding/unfolding of protein A}.
\newblock \emph{Proc Nat Acad Sci USA}. 100:13898--13903.

\bibitem[{Hanggi et~al.(1990)Hanggi, Talkner, and Borkevec}]{Hanggi90}
Hanggi, P., P.~Talkner, and M.~Borkevec. 1990.
\newblock Reaction-rate theory: fifty years after kramers.
\newblock \emph{Rev Mod Phys}. 62:251--341.

\bibitem[{Jackson and Fersht(1991)}]{JacksonSE91i}
Jackson, S.~E., and A.~R. Fersht. 1991.
\newblock Folding of chymotrypsin inhibitor 2. 1. {E}vidence for a two state
  transition.
\newblock \emph{Biochemistry}. 30:10428--10435.

\bibitem[{Klimov and Thirumalai(1997)}]{Klimov97}
Klimov, D.~K., and D.~Thirumalai. 1997.
\newblock Viscosity dependence of the folding rates of proteins.
\newblock \emph{Phys Rev Lett}. 79:317--320.

\bibitem[{Kuhlman et~al.(1997)Kuhlman, Luisi, Evans, and Raleigh}]{KuhlmanB98}
Kuhlman, B., D.~L. Luisi, P.~A. Evans, and D.~P. Raleigh. 1997.
\newblock Global analysis of the effects of temperature and denaturant on the
  folding and unfolding kinetics of the n-terminal domain of the protein l9.
\newblock \emph{J Mol Biol}. 284:1661--1670.

\bibitem[{Lapidus et~al.(2000)Lapidus, Eaton, and Hofrichter}]{LapidusLJ00}
Lapidus, L.~J., W.~A. Eaton, and J.~Hofrichter. 2000.
\newblock Measuring the rate of intramolecular contact formation in
  polypeptides.
\newblock \emph{Proc Nat Acad Sci USA}. 97:7220--7225.

\bibitem[{Leach et~al.(1966)Leach, Nemethy, and Scheraga}]{LeachSJ66}
Leach, S.~J., G.~Nemethy, and H.~A. Scheraga. 1966.
\newblock Computation of sterically allowed conformations of peptides.
\newblock \emph{Biopolymers}. 4:369--407.

\bibitem[{Lindberg et~al.(2002)Lindberg, Tangrot, and Oliveberg}]{LindbergM02}
Lindberg, M., J.~Tangrot, and M.~Oliveberg. 2002.
\newblock Complete change of the protein folding transition state upon circular
  permutation.
\newblock \emph{nsb}. 9:818--822.

\bibitem[{Mines et~al.(1996)Mines, Pascher, Winkler, and Gray}]{Mines96}
Mines, G.~A., T.~Pascher, S.~C. L. J.~R. Winkler, and H.~B. Gray. 1996.
\newblock Cytochrome c folding triggered by electron transfer.
\newblock \emph{Chem. and Biol.} 3:491--497.

\bibitem[{Onuchic et~al.(1997)Onuchic, Luthey-Schulten, and
  Wolynes}]{Onuchic97}
Onuchic, J.~N., Z.~Luthey-Schulten, and P.~G. Wolynes. 1997.
\newblock Theory of protein folding: The energy landscape perspective.
\newblock \emph{Annu Rev Phys Chem}. 48:545--600.

\bibitem[{Otzen and Oliveberg(2004)}]{OtzenDE04}
Otzen, D.~E., and M.~Oliveberg. 2004.
\newblock Correspondence between anomalous m- and $\delta c_p$-values in
  protein folding.
\newblock \emph{Protein Sci.} 13:3253--3263.

\bibitem[{Perl et~al.(2002)Perl, Jacob, Bano, Stupak, Antalik, and
  Schmid}]{PerlD02}
Perl, D., M.~Jacob, M.~Bano, M.~Stupak, M.~Antalik, and F.~X. Schmid. 2002.
\newblock Thermodynamics of a diffusional protein folding reaction.
\newblock \emph{Biophys. Chem.} 96:173--190.

\bibitem[{Plotkin and Onuchic(2000)}]{PlotkinSS00:pnas}
Plotkin, S.~S., and J.~N. Onuchic. 2000.
\newblock Investigation of routes and funnels in protein folding by free energy
  functional methods.
\newblock \emph{Proc Nat Acad Sci USA}. 97:6509--6514.

\bibitem[{Plotkin and Onuchic(2002{\natexlab{a}})}]{PlotkinSS02:quartrev1}
Plotkin, S.~S., and J.~N. Onuchic. 2002{\natexlab{a}}.
\newblock Understanding protein folding with energy landscape theory i: Basic
  concepts.
\newblock \emph{Quart. Rev. Biophys.} 35:111--167.

\bibitem[{Plotkin and Onuchic(2002{\natexlab{b}})}]{PlotkinSS02:quartrev2}
Plotkin, S.~S., and J.~N. Onuchic. 2002{\natexlab{b}}.
\newblock Understanding protein folding with energy landscape theory ii:
  Quantitative aspects.
\newblock \emph{Quart. Rev. Biophys.} 35:205--286.

\bibitem[{Portman et~al.(2001)Portman, Takada, and Wolynes}]{PortmanJJ01:jcp}
Portman, J.~J., S.~Takada, and P.~G. Wolynes. 2001.
\newblock Microscopic theory of protein folding rates. {II.} {Local} reaction
  coordinates and chain dynamics.
\newblock \emph{J. Chem. Phys.} 114:5082--5096.

\bibitem[{Scalley and Baker(1997)}]{ScalleyM97}
Scalley, M., and D.~Baker. 1997.
\newblock Protein folding kinetics exhibit an arrhenius temperature dependence
  when corrected for the temperature dependence of protein stability.
\newblock \emph{Proc Nat Acad Sci USA}. 94:10636--10640.

\bibitem[{Schindler and Schmid(1996)}]{SchindlerT96}
Schindler, T., and F.~X. Schmid. 1996.
\newblock Thermodynamic properties of an extremely rapid protein folding
  reaction.
\newblock \emph{Biochemistry}. 35:16833--16842.

\bibitem[{Shea and {Brooks~III}(2001)}]{SheaJE01}
Shea, J.~E., and C.~L. {Brooks~III}. 2001.
\newblock From folding theories to folding proteins: A review and assessment of
  simulation studies of protein folding and unfolding.
\newblock \emph{Annual Review of Physical Chemistry}. 52:499--535.

\bibitem[{Shoemaker et~al.(1999)Shoemaker, Wang, and Wolynes}]{ShoemakerWang99}
Shoemaker, B.~A., J.~Wang, and P.~G. Wolynes. 1999.
\newblock Exploring structures in protein folding funnels with free energy
  functinals: The transition state ensemble.
\newblock \emph{J Mol Biol}. 287:675--694.

\bibitem[{Socci et~al.(1996)Socci, Onuchic, and Wolynes}]{SocciND96:jcp}
Socci, N.~D., J.~N. Onuchic, and P.~G. Wolynes. 1996.
\newblock Diffusive dynamics of the reaction coordinate for protein folding
  funnels.
\newblock \emph{J Chem Phys}. 104:5860--5868.

\bibitem[{Wang et~al.(1997)Wang, Plotkin, and Wolynes}]{WangPlot97}
Wang, J., S.~S. Plotkin, and P.~G. Wolynes. 1997.
\newblock Configurational diffusion on a locally connected correlated energy
  landscape; application to finite, random heteropolymers.
\newblock \emph{J. Phys. I France}. 7:395--421.

\bibitem[{Zhou et~al.(2003)Zhou, Zhang, Stell, and Wang}]{ZhouY03}
Zhou, Y., C.~Zhang, G.~Stell, and J.~Wang. 2003.
\newblock Temperature dependence of the distribution of the first passage time:
  Results from discontinuous molecular dynamics simulations of an all-atom
  model of the second beta-hairpin fragment of protein g.
\newblock \emph{J Am Chem Soc}. 125:6300--6305.

\end{thebibliography}
%\newpage

\vspace{0.5in}
\centerline{FIGURE CAPTIONS}

\vspace{0.6in}

\begin{figure}
\includegraphics[height=5cm]{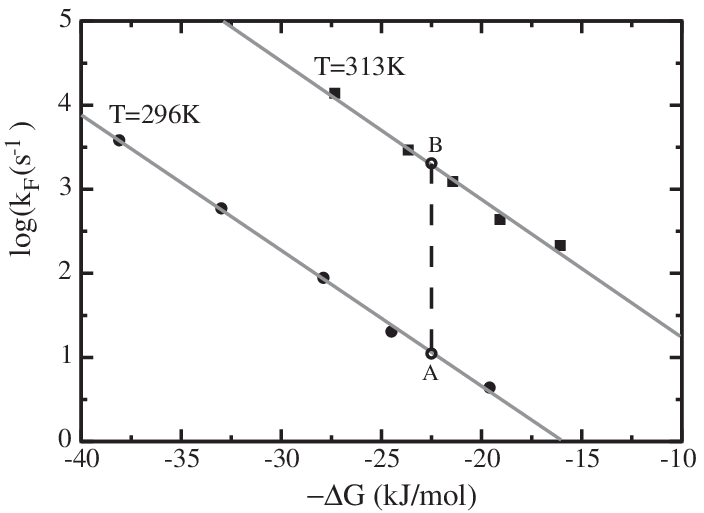}
\caption{\label{fig:gray}
}
\end{figure}

FIGURE \ref{fig:gray}: 
Logarithm of the rate {\it vs.} (minus) native stability for horse
Cytochrome~C, at two temperatures. The plots are well fit by straight
line functions that are used in the analysis of the text. Adapted from
Mines {\it et. al.}~\citep{Mines96}.

\vspace{0.6in}

\begin{figure}
\includegraphics[height=5cm]{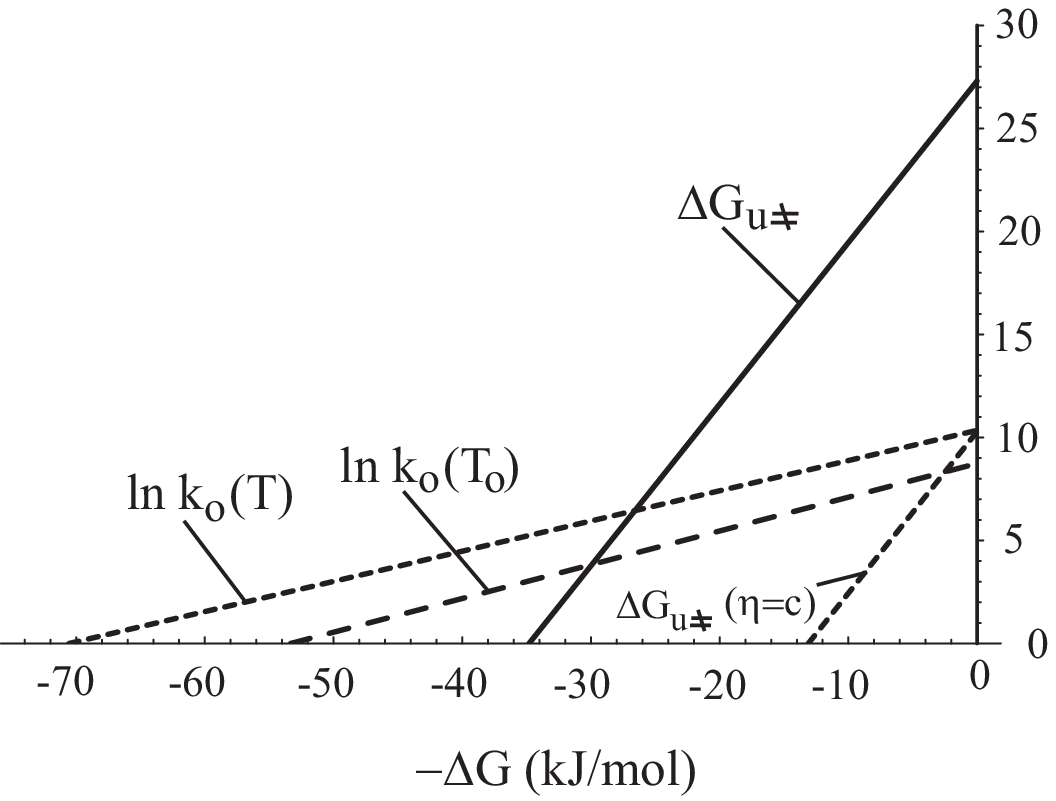}
\caption{\label{fig:grayDG}
}
\end{figure}

FIGURE \ref{fig:grayDG}: 
Barrier height $\DGud$ and prefactors $k_o$ at two temperatures, as obtained from the REM
approximation (see text) are plotted as a function of minus stability,
for cytochrome~C. The wild type protein has a stability of $\DG
\approx 74$ kJ/mol. Numerical values are given in table~I. 
Prefactor attempt rates are in $\mbox{s}^{-1}$, and barrier heights
are in kJ/mol. The short dashed line gives the barrier for a
temperature-independent solvent viscosity.

\vspace{0.6in}

\begin{figure}
\includegraphics[height=10cm]{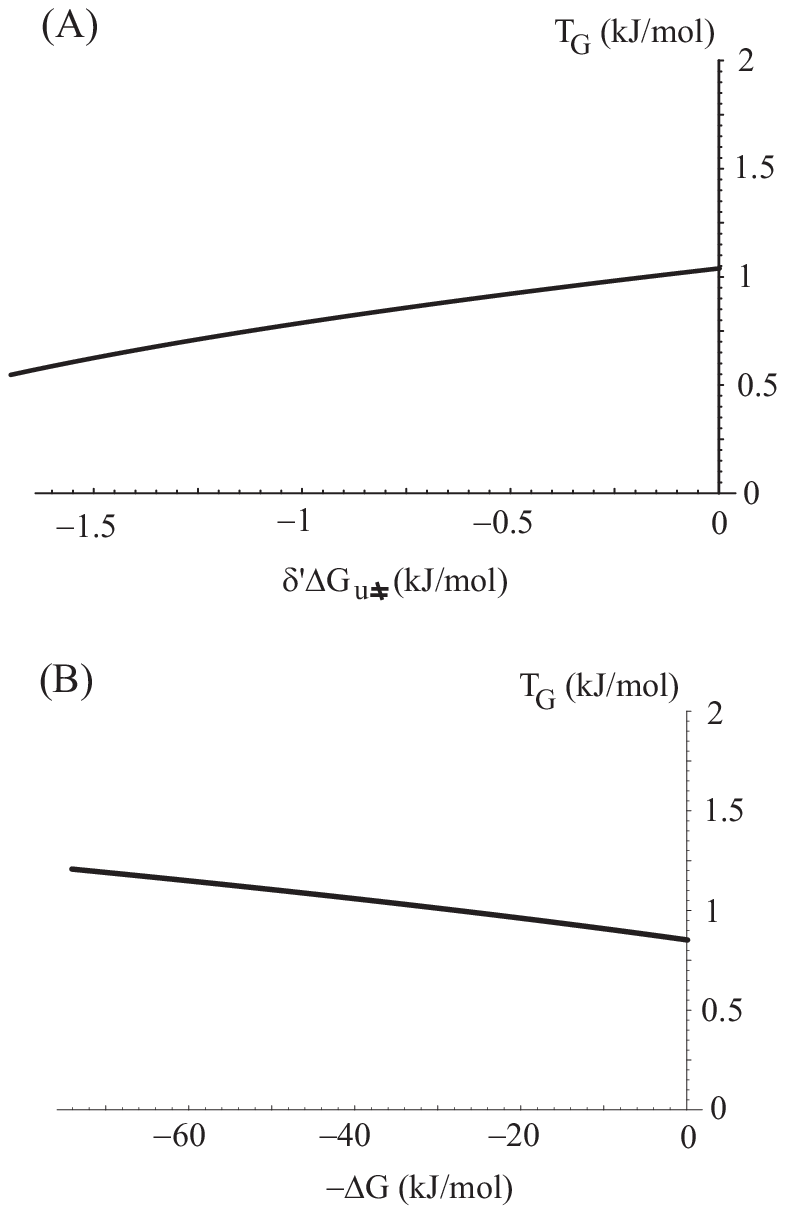}
\caption{\label{fig:tg}
}
\end{figure}

FIGURE \ref{fig:tg}:
{\bf (A)} The temperature $\Tg$ that emerges from the REM analysis for cyt-C
(see text and eq.~\ref{tg}) varies only moderately with
barrier height change at constant stability, $\d' \DGud$ (the value of
which is not
known for this protein). 
For this plot the stability is set to midway between zero and the
stability of the wildtype ($37$ kJ/mol). {\bf (B)} $\Tg$ also changes little as 
native stability $\DG$ is varied (for this plot $\d' \DGud =0$).

\vspace{0.6in}

\begin{figure}
\includegraphics[height=5cm]{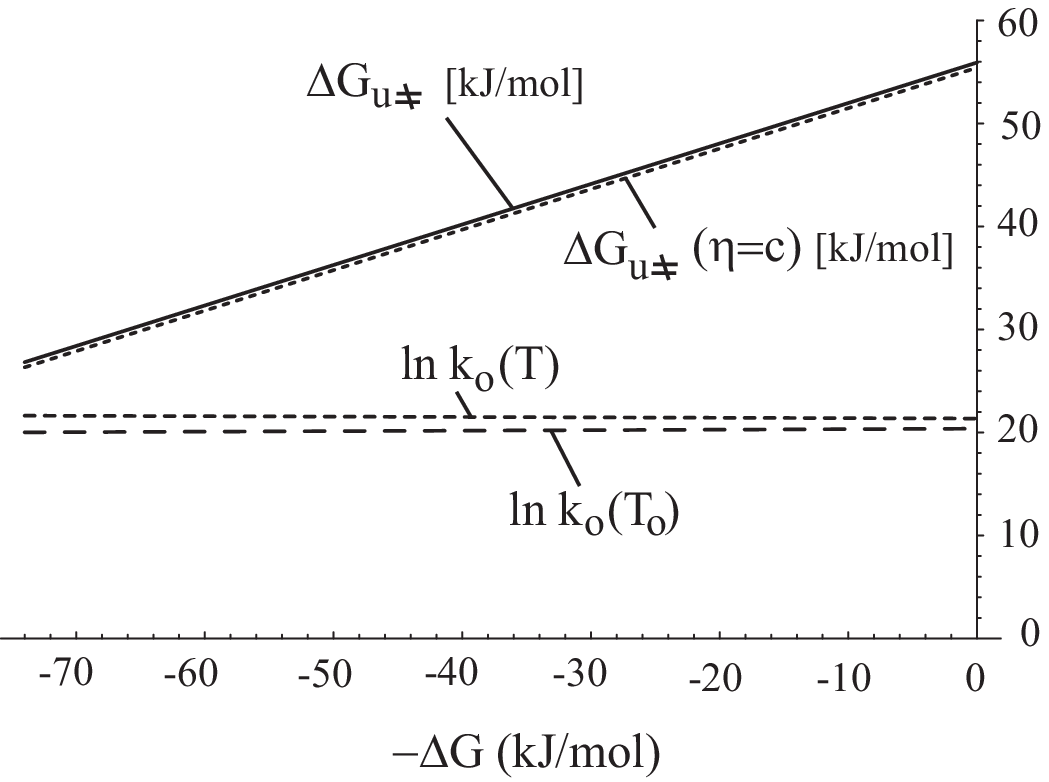}
\caption{\label{fig:graygrem}
}
\end{figure}

FIGURE \ref{fig:graygrem}:
Barrier heights and prefactors as obtained from the correlated
landscape analysis (see text), plotted as a function of minus native
stability for h cytC. Numerical values are given in table~I. Prefactor 
attempt rates are in $\mbox{s}^{-1}$, and barrier heights
are in kJ/mol. The dotted line gives the barrier for a
temperature-independent solvent viscosity. Note prefactors are roughly
constant and solvent viscosity plays a minor role.

\vspace{0.6in}

\begin{figure}
\includegraphics[height=5cm]{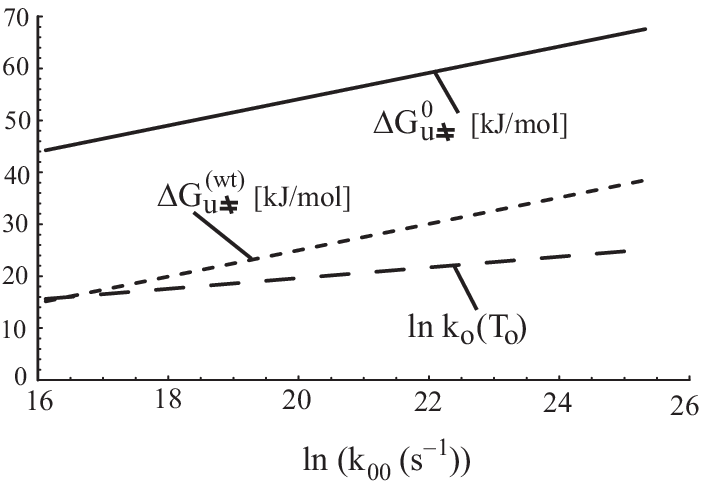}
\caption{\label{lnkoofig}
}
\end{figure}

FIGURE \ref{lnkoofig}:
Barrier heights and prefactors extracted from the recipe for the correlated energy
landscape (see text) increase as the bare reconfiguration rate (defined
in~\ref{lnkogrem} and~\ref{lnkgrem}) increases. 
The increase is linear. $\DGud^o$ is the barrier at the transition
midpoint,  $\DGud^{(wt)}$ is the barrier at the stability of the
wild-type protein, and $k_o(T_o)$ is the prefactor at temperature
$T_o$ in $\mbox{s}^{-1}$.

%\newpage

%\newpage

%\newpage

%\newpage

%\newpage

\vspace{10cm}
%\newpage

%\newpage

\pagestyle{empty}
%\baselineskip=0.2cm
%\vspace{-1.0cm}
\begin{table}
\label{tab:1}
\caption{
{\bf Thermodynamic and kinetic parameters for proteins studied}
}
\begin{ruledtabular}
%\begin{tabular}{cccccccccccccccc}
\begin{tabular}{l|llll|llll|llll}
& \multicolumn{4}{c}{}
& \multicolumn{4}{c}{Correlated Model}
%\vline
&  \multicolumn{4}{c}{Random Energy Model}
  \\
Proteins \footnotemark[1] & $T_o$ & $T$ & $\DG^{\mbox{\tiny{WT}}}$
  \footnotemark[2] & $\d'\DGud$ & $\DGud^o$
\footnotemark[3]& $\DGud^{wt}$  \footnotemark[4]&
$k_o (T_o)$ \footnotemark[3] & $\Tg^{\left( m\right)}$
  \footnotemark[5] &
$\DGud^o$\footnotemark[3] & $\DGud^{wt}$ \footnotemark[4]& $k_o
(T_o)$ \footnotemark[3] & $\Tg^{\mbox{\tiny REM}}$ \footnotemark[6] \\ \hline 

cyt C   &
2.46 & 2.60  & -74 & 0 & 
56 &  27  & 5$\times 10^{8}$ & 1.2 & 
27 & 0 &  6 $\times 10^3$ &  1.0   \\

NTL9 & 
2.48  & 2.59 & -19 & -1.0  & 
47 & 35 & ($6\times 10^{9}$) & 1.7  & 
50 & 33 & ($10^{10}$) & 0 (0.4) \footnotemark[7] \\

S6 &
2.48 & 2.56 & -31 & -1.4  & 
61 & 39 & $9\times 10^{8}$ & 1.2  & 
78 & 21 & ($10^{12}$) & 0 (0.7) \footnotemark[7] \\

PTL  & 
2.34  & 2.43  & -22 & -0.8  & 
58 & 45  & 1$\times 10^{9}$ & 1.1 & 
56 & 27 & 3$\times 10^{8}$ & 0.5  \\

cspB  & 
2.38 & 2.44 & -9 & -0.8 & 
7 & 0 & $10^2$ & 1.9 \footnotemark[7]  & 
24 & 20 & 2$\times 10^{5}$  & 0.7

\end{tabular}
\footnotetext[1]{Sources for experimental data: cyt C~\citep{Mines96},
  NTL9~\citep{KuhlmanB98}, S6~\citep{OtzenDE04}, PTL~\citep{ScalleyM97},
  cspB~\citep{SchindlerT96}. All temperatures and energies are in
  kJ/mol. All rates are in~$\mbox{s}^{-1}$.}
\footnotetext[2]{Stability of the wild type protein.}
\footnotetext[3]{At the transition midpoint where $\DG =0$.}
\footnotetext[4]{At the stability of the wild-type protein, where
  $c=0$. If the barrier vanished at stabilities below the wild type, the
  barrier value was simply taken as zero.}
\footnotetext[5]{Value of $\Tg$ that gives a slope of barrier height
  vs. stability equivalent to the experimental value of $\mud/m$.}
\footnotetext[6]{Value of $\Tg$ using the REM approximation for
  rates, taken at a  stability of about $1/2$ of the wild-type
  protein. }
\footnotetext[7]{See text for explanation and comments.}
%\footnotetext[8]{
%The REM value of $\Tg$ is used here, when a value
%  closer to $\Tg^{(m)}$ is used, the corresponding values are 
%$\DGud^o = 6$ kJ/mol, $\DGud^{wt} = 0$ kJ/mol, $k_o(T_o) = 10^3
%  \mbox{s}^{-1}$  (see text).}
\end{ruledtabular}
\end{table}

\end{document}